\def\PR{{Phys.~Rev.~}}
\def\PRL{{ Phys.~Rev.~Lett.~}}
\def\PRB{{ Phys.~Rev.~B~}}
\def\PRA{{ Phys.~Rev.~A~}}

\def\JCP{J.~Chem. Phys.~}

\def\expt10{J.~Paier, M.~Marsman, K.~Hummer, G.~Kresse, I.~C.~Gerber and J.~G.~\'Angy\'an}

\def\CdO186{P.~D.~C.~King, T.~D.~Veal, A.~Schleife, J.~Z\'u$\tilde{n}$iga-P\'erez, B.~Martel, P.~H.~Jefferson, F.~Fuchs, V.~Mu$\tilde{n}$oz-Sanjos\'e, F.~ Bechstedt and C.~F.~McConville}
\def\etal{{\it et al.}}

\def\be{\begin {equation}}
\def\ee{\end {equation}}
\def\ber{\begin {eqnarray}} 
\def\eer{\end {eqnarray}}
\def\bers{\begin {eqnarray*}}
\def\eers{\end {eqnarray*}}
\def\eq{\ =\ }

\newcommand{\angstrom}{\AA}
\makeatletter

\newcommand{\Rmnum}[1]{\expandafter\@slowromancap\romannumeral #1@}
\makeatother

\makeatletter
\newcommand*\env@matrix[1][*\c@MaxMatrixCols c]{%
  \hskip -\arraycolsep
  \let\@ifnextchar\new@ifnextchar
  \array{#1}}
\makeatother

\documentclass[aps,prb,showpacs,superscriptaddress,a4paper,twocolumn]{revtex4-1}

\usepackage{amsmath}
\usepackage{float}
\usepackage{subfigure}
\usepackage{color}

\usepackage{color, colortbl}
\definecolor{Gray}{gray}{0.75}
\definecolor{LightCyan}{rgb}{0.50,1,1}

\usepackage[section]{placeins}
\usepackage{amsfonts}
\usepackage{amssymb}
\usepackage{graphicx}

\begin {document}

\author{Prashant Singh}\email{prashant@ameslab.gov}
\affiliation{Ames Laboratory, U.S. Department of Energy, Iowa State University, Ames, Iowa 50011-3020, USA}
\author{Manoj K Harbola}\email{mkh@iitk.ac.in}
\affiliation{Department of Physics, Indian Institute of Technology, Kanpur, 208016, India}
\author{M Hemanadhan}
\altaffiliation[Present address: ]{D\'{e}partement de Chimie Mol\'{e}culaire, Universit\'{e} Joseph Fourier, Grenoble Cedex 9, 38041, France}
\affiliation{Department of Physics, Indian Institute of Technology, Kanpur, 208016, India}
\author{Abhijit Mookerjee}\email{abhijit.mookerjee61@gmail.com}
\affiliation{S. N. Bose National Centre for Basic Sciences, Salt Lake City, Kolkata, 700098, India}
\author{D. D. Johnson}\email{ddj@ameslab.gov}
\affiliation{Ames Laboratory, U.S. Department of Energy, Iowa State University, Ames, Iowa 50011-3020, USA}
\affiliation{Materials Science \& Engineering, Iowa State University, Ames, Iowa 50011-2300, USA} 

\title{Better Band Gaps with  Asymptotically-Corrected Local-Exchange Potentials}

\begin{abstract}
We formulate a spin-polarized van Leeuwen and Baerends (vLB)  correction  to the local density approximation (LDA) exchange potential $\left[\PRA {\bf 49}, 2421~(1994)\right]$ that enforces the ionization potential (IP) theorem following Stein et al.~$\left[\PRL {\bf 105}, 266802~(2010)\right]$. For electronic-structure problems, the vLB-correction replicates the behavior of exact-exchange potentials, with improved scaling and well-behaved  asymptotics, but with the computational cost of semi-local functionals. The vLB+IP corrections produces large improvement in the eigenvalues over that from LDA due to correct asymptotic behavior and atomic shell structures, as shown on rare-gas, alkaline-earth, zinc-based oxides, alkali-halides, sulphides, and nitrides. In half-Heusler alloys, this asymptotically-corrected LDA reproduces the  spin-polarized properties correctly, including magnetism and half-metallicity. We also considered finite-sized systems [e.g., ringed boron-nitirde (B$_{12}$N$_{12}$) and graphene (C$_{24}$)] to emphasize the wide applicability of the method.
\end{abstract}
\date{\today}

\maketitle

\section{Introduction}
Density functional theory (DFT)\cite{KK2008,VS2004,HK1964,KS1965} is the most widely used methods to explore electronic binding in materials, and uses approximate functionals for exchange-correlation (XC) energy calculation. Foremost among them is the local density approximation (LDA),\cite{PW1992} which,  over the years, has been improved substantially by developing generalized gradient-corrected approximation (GGA) functionals.\cite{PCVJPSF1992,PRCVSCZB2008}  While these functionals have been quite successful in predicting a large number of properties and are used widely for large systems (due to their computational efficiency and reasonable accuracy), almost all semi-local functionals fail measurably in predicting the correct band gaps. Attempts have been made to improve band-gap prediction using semi-local approaches, such as Self-Interaction Correction (SIC) method,\cite{JP1979,PZ1981,PC1988}  and DFT plus Hubbard corrections (DFT+U).\cite{AZA1991} Other commonly used approaches  are GW approximation,\cite{ASG2000,FSK2004,SKF2006,CSK2007,SK2007,SMK2007} and hybrid functionals.\cite{HPSM2005,PMHKGA2006} However, they are often limited to small system sizes due to their large computational demand. 
One significant reason for the partial failure of semi-local XC potentials is their inability to describe the correct asymptotic behavior,\cite{PZ1981} leading to qualitatively incorrect results for properties sensitive to the asymptote, e.g., the fundamental gap\cite{PPLB1982,SS1983,PL1983, DJT2003} and ionization potential.\cite{CJCS1998} 

{\par}The poor band-gap predictions in solids using semi-local functionals is understood to arise from the failure to describe correctly the discontinuous jump $\Delta_{xc}$ (a constant) in Kohn-Sham (KS) potential as the electron number crosses an integer value.\cite{PPLB1982,SS1983,PL1983,DJT2003,KK2014} From the continuity of KS orbitals across this jump,\cite{PPLB1982,SS1983,PL1983,DJT2003,KK2014} it directly follows that  $\Delta_{xc}= E_{g} - E_{g}^{KS}$, where the fundamental gap ($E_{g} \equiv  I - A$) is the difference between the ionization potential, $I$, and the electron affinity, $A$, while the  KS gap  ($E_{g}^{KS} \equiv \epsilon_{LU} - \epsilon_{HO} $) is the difference between the lowest-unoccupied (LU)  and highest-occupied (HO)  eigenvalues.
Any error in $\Delta_{xc}$ (including asymptotic behavior) leads to large deviation from the ionization potential (IP) theorem and underestimation of $E_{g}$.\cite{PPLB1982,SS1983,PL1983,DJT2003,KK2014}  
Exact-exchange (EXX) functionals possess $\Delta_{xc}$ by construction.\cite{KK2008,GKG1997,ED2011} Kotani implemented EXX in the KS framework for solids and showed substantial improvement in $E_{g}$ and asymptotic behavior of the potential.\cite{TK1995} Following this, many attempts were made to mimic EXX behavior with semi-local functionals.\cite{BJ2006,TBS2007,TB2009,KOER2010,AK2013} The van Leeuwen--Baerends (vLB) correction to LDA-exchange used for atoms is one such approach; orbital eigenenergy differences calculated using this approach are close to atomic excitation energies.\cite{VLB,UWW1988,UG1994}

{\par}Recently, Kraisler and Kronik\cite{KK2014} showed that all XC functionals (local, semi-local,  and non-local) generally possess a non-zero  $\Delta_{xc}$, and addressed the estimation of $E_{g}$  within approximate density functionals from ensemble considerations. 
For finite systems (including small periodic cells) they showed that $\Delta_{xc}$ from semi-local functionals  dramatically improves the predicted $E_{g}$, even for LDA; however, as the system is extended (i.e., large supercells), the ensemble correction for LDA vanishes. The main difficulty arises because the HO and LU orbitals are delocalized whereas the XC kernel is very localized in semi-local cases. To avoid addressing the highly non-local kernels, an alternative is to localize the HO and LU orbitals, such as by dielectric screening,\cite{CC2010} SIC, or use of small cells (an uncontrolled localization). 
An alternative is to impose the asymptotic behavior in a solid locally as part of the electronic-structure method, as easily implemented in site-centered basis-set methods using vLB-correction for solids.

{\par}Using the Harbola-Sahni (HS) exchange-potential\cite{HS1989} and vLB correction,\cite{VLB} we have previously established  that correcting asymptotic limits of exchange-potential leads to significant improvement in semiconductor band gaps.\cite{PHA2013,PHA2015} Here, the vLB-corrected LDA\cite{VLB} is much easier to apply because it can be  written in terms of the system density. The vLB-correction\cite{VLB} is constructed along the lines of the Becke-functional, \cite{ADB1988} and makes the XC-potential go asymptotically as $-1/r$ ``far'' from the atom-center but still inside the crystal, which must be the local interstitial region surrounding each atom in the solid, and which also defines the local crystal potential zero (to solve accurately the  microscopic electrostatics). This approach may be connected (Section~\ref{vLBsec}) to range-separated functionals that invoke the asymptotic condition $-1/(\epsilon r)$ involving the dielectric function ($\epsilon$) of the solid. \cite{GAMK2007,ASJBNK2013,AJSNK2015} 

{\par}In this paper, we implement a spin-polarized version of vLB-correction to the short-range part of LDA exchange, which is similar to the modified-LDA,\cite{TB2009} rather than the full range addressed in range-separated hybrid functionals.\cite{GAMK2007,ASJBNK2013,AJSNK2015} We apply it to a wide range of materials with varying crystal structures, e.g., rare-gas solids, nitrides, oxides, sulphides, ternary half-Heusler alloys and some finite size systems. In particular, we address wurtzite-ZnO, spin-polarized, half-Heusler (C1$_{b}$) FeMnSb, and two-dimensional boron-nitride (B$_{12}$N$_{12}$). Our spin-polarized vLB-corrected LDA corrects both extreme limits of the potential, i.e., $r\rightarrow{0}$ and $r\rightarrow{\infty}$, and uses the optimized vLB parameter to obey the ionization-potential (IP) theorem\cite{PPLB1982,SS1983,AB1985} for isolated atoms and diatoms;  these combined corrections given substantial improvement of semiconductor band gaps over LDA, and are similar to EXX results.

\section{BACKGROUND}
\subsection*{Ionization Potential Theorem}
Perdew \etal\cite{PPLB1982} has shown that for the case of exact KS theory in DFT, the highest-occupied KS eigenvalue is equal and opposite to the ionization potential. Stein \etal\cite{SEKB2010} has followed the queue from the IP-theorem\cite{PPLB1982,SS1983,AB1985} to determine the optimal value of system independent parameter $\gamma$\cite{DFPLMC2010} used to evaluate XC-part in their calculation and successfully extended  the quantitative usage of DFT for calculating fundamental gaps in finite and bulk systems, a largely unsubscribed area.\cite{SEKB2010,ASJBNK2013,AJSNK2015} 

{\par}Hemandhan \etal\cite{HSH2014} has combined the idea of Stein \etal\cite{SEKB2010} with IP-theorem\cite{PPLB1982,SS1983,AB1985} to determine the optimize value of $\beta$ used in vLB correction-term.\cite{VLB} The vLB-correction to LDA exchange implemented with Vosko-Wilk-Nusair parametrized correlation\cite{VWN1980} to calculate excitation energies of atoms and diatoms following IP-theorem. To make the optimal choice of $\beta$ and determining the highest-occupied molecular orbital (HOMO) energies, we enforce Koopmans theorem, i.e., $\beta$ varied until the HOMO eigenvalue $\epsilon_{max}$ and ionization energy matches,\cite{HSH2014} i.e.,
\begin{equation}\label{IPT}
\epsilon_{max}^{\beta} = E(N,\beta) - E(N-1,\beta)=-I^{\beta}(N)  .
\end{equation}
here, N is the number of electrons in system, $I^{\beta}(N)$ is the energy difference between the ground state energies, of the $N$ and the $(N-1)$ electron system per $\beta$, i.e., ionization potential.

\subsection*{van Leeuwen--Baerends correction}\label{vLBsec}
{\par}To apply KS-DFT to  a system of interest, we need to approximate XC-potential in evaluating the effective potential as accurately as possible. This potential is largely evaluated by taking derivative of XC-energy functional but system density can also be used to model them directly. Recent reports,\cite{AKK2008, KAK2009, KOER2010, KTB2011} indicate that direct approximation approach to the KS-potentials can be a promising route for accurate prediction of static electric polarizabilities, band gaps, and other properties. One such example is the correction introduced by van Leeuwen and Baerends,\cite{VLB} model potential to the exchange part of XC-potential. We employ vLB-correction to LDA exchange as
\begin{equation}
V_{xc,\sigma}^{model}({\bf r}) = \left[V_{x,\sigma}({\bf r}) + V_{x,\sigma}^{vLB}({\bf r})\right] + V_{c,\sigma}({\bf r})
\label{MODEL}
\end{equation}
where $V_{x,\sigma}(\bf r)$[$V_{c,\sigma}({\bf r})$] is the standard LDA exchange[correlation] potential\cite{PW1992} and $V_{x,\sigma}^{vLB}({\bf r})$ is correction to LDA exchange.\cite{VLB} The suffix $\sigma$ represents the spin degree of freedom. Here, $V_{x,\sigma}^{vLB}({\bf r})$ is 
\begin{equation}
V_{x,\sigma}^{vLB} ({\bf r}) = -\beta\rho_{\sigma}^{1/3}\frac{x_{\sigma}^{2}}{1+3{\beta x_{\sigma}}{\sinh}^{-1}(x_{\sigma})},
\label{VLB}
\end{equation}
where $\beta=0.05$ was used in the original  formulation.\cite{VLB} The variable $x=|\nabla\rho({\bf r})|/{\rho^{4/3}({\bf r})}$ signifies the change in mean electronic distance provided density is a slowly varying function in given region with strong dependence on gradient of local radius of the atomic sphere $R_{ASA}$.

{\par} The effective Kohn-Sham potential using Eq.~\ref{VLB} is
\begin{eqnarray}\label{VEFF}
V_{\rm eff}({\bf r}) = V_{ext}({\bf r}) &+& V_{H}({\bf r})  \nonumber  \\
              &+& [V_{x,\sigma}({\bf r})+V_{x,\sigma}^{vLB}({\bf r})+V_{c,\sigma}({\bf r})]\phantom{XIX}  ,
\end{eqnarray}
where the potential contributions are external V$_{ext}({\bf r})$, electronic Hartree V$_{H}({\bf r})$, LDA exchange V$_{x,\sigma}({\bf r})$,  LDA correlation V$_{c,\sigma}({\bf r})$, and is the spin-polarized  vLB-exchange V$_{x,\sigma}^{vLB}({\bf r})$. The optimized $\beta$, calculated from IP theorem for atoms and diatoms, helps in calculating accurate densities in solids. This provides us a general procedure to construct the KS XC-potential from a given electron density and produces fairly good asymptotic behavior along with fulfilling the requirements of EXX-potential.\cite{VLB}

{\par} The iterative Kohn-Sham scheme  now has the effective potential constructed using new electronic density term
\ber \left\{-\frac{1}{2} \nabla^2   + (V_{\rm eff}({\bf r}) - V_o)  \right\} \phi_{i,\sigma}({\bf r}) \nonumber   \\ 
     = (\epsilon_{i,\sigma} - V_o)\phi_{i,\sigma}({\bf r}) .\phantom{XX XX}
\label{KS1}\eer
In Eq.~\ref{KS1}, we call-out the potential zero $V_o$ directly, which is arbitrary in ``exact'' full-potential linear augmented plane wave (FLAPW), but not so in approximate methods.
We employ Eq.~\ref{KS1} in tight-binding, linear muffin-tin orbital method (TB-LMTO) within the atomic sphere approximation (ASA)\cite{TBLMTO}  and obtain the self-consistent solution of the single-particle Schr\"odinger equation for the vLB-corrected effective potential.

{\par} Many site-centered electronic-structure methods, such as LMTO, KKR, and FLAPW utilize a spherical-harmonic basis within a specified radius and then handle interstitial regions between atoms in varying levels of accuracy -- FLAPW  is considered ``exact'' by using thousands of plane waves to represent unit cell and interstitial. Within the ASA, the atomic spheres are simply increased to conserve cell volume while large interstitial voids are represented by non-overlapping empty spheres (ESs). The ES contain no ion cores but associated charge reflects interstitial potential. Then, all potentials and eigen-energies are chosen relative to a suitable $V_o$, which can be set variationally so that the ASA dispersion approach that of FLAPW,\cite{Alam.PRB.85.144202.2012}  and establishes the ``free-electron'' like behavior inside the crystal (and defines the crystal momentum).

{\par}Importantly, LMTO methods, via the tail-cancellation theorem,\cite{Skriver1983} or, equivalently, KKR methods, through the calculation of the single-site scatterers, permit the vLB-correction to be imposed locally for all sites in a multisite (finite or infinite) structure to solve for the collective behavior. That is, the interstitial region within the crystal is effectively the asymptotic region for site-centered methods, where the vLB-correction is set at the ASA boundary. Hence, as is typical, we solve the microscopic electrostatic potential inside the crystal, without direct reference to the atomic zero far outside the crystal.

{\par} To connect our results to those of methods that use screened, range-separated hybrid functionals implemented in plane-wave methods, we note that the global reference is set to atomic zero (far from the atom or far outside the crystal), which requires a dielectric function to solve the macroscopic (long-ranged) electrostatics, as done, for example, by Kronik et al.\cite{GAMK2007,ASJBNK2013,AJSNK2015} In this approach, to set the proper boundary conditions,  two of the range-partition variables must obey the sum rule $\alpha + \beta = 1/\epsilon$, where $\epsilon$ is the static dielectric function ($\epsilon=1$ for an atom in vacuum and $1 \le \epsilon \le \infty$ for a crystal). Within our site-centered basis method with potential reference $V_o$, there is a difference to the atomic zero due to the work function, i.e.,  $W = -E_F - e\phi$, where $\phi \propto \epsilon$ for systems with a gap. Hence, the present theory should (and does) reproduce the band gap results of Kronik et al.,\cite{GAMK2007,ASJBNK2013,AJSNK2015} without the need to calculate the static dielectric as input. Of course, we can also calculate the $W$ to establish our results relative to atomic zero, for which the long-range Fock and semi-local exchange play a role.

\section{COMPUTATIONAL DETAILS}
{\par} Core states are treated as atomic-like in a frozen-core approximation and energetically higher-lying valence states are addressed in the self-consistent calculations of the effective crystal potential, which is constructed by overlapping Wigner-Seitz spheres for each atom in the unit cell. A two-fold criteria for generating crystal potential, on the same footings of TB-LMTO-ASA, has been used: $\left(a\right)$ use of trial wave function, i.e., linear combinations of basis functions like plane waves in the nearly free-electron method, and $\left(b\right)$ use of matching condition for partial waves at the sphere boundary.\cite{TBLMTO,OKA1975,AJS1986}

{\par}All spin-polarized vLB-correction calculations were done self-consistently and non-relativistically for given experimental geometry until the ``averaged relative error'' between successive iterations reaches  10$^{-5}$ for charge density and 10$^{-4}$ for energy. To facilitate convergence, we have used Anderson mixing. The k-space integration is done using the tetrahedron method with divisions of~$12\times12\times12$ for cubic and $12\times12\times6$  for non-cubic cells along the three primitive reciprocal translation vectors.

{\par} Inside the atomic spheres, the Kohn-Sham potential is obtained by using the LDA correlation parameterized by van Barth and Hedin\cite{vBH1972} with the corrected EX potential given by Eq.~\ref{VEFF}, matched at the ASA radii. Following TB-LMTO-ASA requirements,\cite{TBLMTO} the open-shell semiconductors structures are filled with ESs for an improved basis.  In empty spheres, given that absence of a core makes the electron gas reasonably homogeneous and small, the exchange contribution from empty spheres is very small and the correction to exchange is even smaller; hence, we use LDA-XC in empty spheres, and implement the vLB-correction in atomic spheres only.

{\par}The dependence of dimensionless parameter $x$ present in Eq.~{\ref{VLB}} on $R_{ASA}$ is very clear from the expression,~so appropriate choice of $R_{ASA}$ is crucial. In all calculations,~we chose $R_{ASA}$ by $\pm$5-10\% from default values to control the overlapping of atomic spheres and empty spheres to reduce the loss of electrons into the (unrepresented) interstitial for open-shell structures, e.g. semiconductors.

{\par} In Section~\ref{rdsec}, we have shown  that vLB-corrected LDA provides an accurate band structure of semiconductor and insulators, but, due to the absence of exchange functional E$_{x}$ such that V$_{x,\sigma}^{vLB}=\delta{E_{x}}/\delta\rho_{\sigma}$, we recommend the use of existing semi-local functionals (e.g., LDA or GGA) for structural properties, and then the vLB-corrected LDA to obtain the band structure.

\section{Results and discussion}\label{rdsec}
Recently, Hemanadhan \etal\cite{HSH2014} discussed extensively that the optimal tuning of the parameter $\beta$ is crucial for achieving an accurate description of IP-theorem for atoms. However, the tuning procedure is challenging in solids because ionization potential and electron affinity need to be calculated from total energy differences, a problematic procedure for periodic systems.\cite{VENKB2014} To achieve this goal, we set two step criteria, first, we tuned the parameter $\beta$ both for atom and diatoms to an optimal value using IP-theorem, e.g., if we choose diatom LiF which has experimental ionization potential of 11.50 eV. Tuning $\beta$ to 0.048 from atomic calculations gives $\epsilon^{\beta}_{max}=-I^{\beta}=12.17 eV$ just 5\% from experimental observation. Secondly, we use the optimal value of $\beta=0.048$ in vLB-correction term to LDA-exchange in Eq.~\ref{VLB}. This way, the calculated band gap of LiF using our self-consistent optimized vLB-TB-LMTO-ASA approach is 12.61 eV which compares well with the calculated number 12.60 eV using range-separated hybrid functional by Refaely-Abramson\etal \cite{AJSNK2015} and experimental value of 13.60 eV.\cite{POL1956}

\begin {table}[b]
\caption{Band gaps calculated with vLB-corrected potential of A1(fcc) and B1(Rocksalt) systems at values of $\beta$ satisfying the  IP theorem. We compared to results from experiments,\cite{POL1956,expt4,expt7,expt10,expt14,expt18,expt23,expt24,expt25,expt26,ExptRG,GaS} HS-EX, LDA, QPC\cite{QPC} and MBJ-LDA.\cite{TB2009}}
\begin {tabular}{cccccccccccc}\hline \hline
System&& \multicolumn{7}{c}{Band Gap (eV)}\\ 
                    &$\beta$&&  vLB   & Expt.   && HS-EX  &  LDA  &&  QPC  & MBJ  \\ \hline     
Ne  (A1) & 	 0.082	 &&23.64  & 20.75 && 22.07 & 11.39 && 16.55 & 22.72   \\
               &    0.05    && 23.02 & &&  & && &  \\
Ar  (A1)  & 	 0.058	 && 12.76 & 14.32 && 11.29 &  8.09  && 11.95 & 13.91   \\
               &   0.05     && 12.46 & &&  & && &  \\
Kr  (A1)  & 	 0.044	 && 10.91 & 11.40 &&  9.10  &  6.76  &&  9.98  & 10.83   \\
               &   0.05     && 10.61 & &&  & && &  \\
Xe  (A1)  & 	 0.040	 &&  8.61  &  9.15  &&  6.63  &  5.56  &&  8.23  &  8.52   \\
               &   0.05     && 8.35   & &&  & && &  \\
MgO (B1)& 	 0.070	 &&  6.94  &  7.78  &&  6.23  &  4.94  &&  ---      &  7.17   \\
               &   0.05     && 5.94   & &&  & && &  \\
CaO (B1)& 	 0.070	 && 7.15   &  7.09  &&  7.29  &  3.36  &&  ---     &  ---    \\
               &   0.05     && 6.07   & &&  & && &  \\
LiF (B1)  & 	 0.048	 && 12.61 & 13.60 &&  9.52  &  8.94  &&  ---     & 12.94  \\
              &    0.05     && 12.31 & &&  & && &  \\         
LiCl(B1)  & 	 0.048	 &&  7.84  &  9.40  &&  6.50  &  6.06  &&  ---     &  8.64   \\
               &   0.05     && 7.85   & &&  & && &  \\
\hline \hline
\end {tabular}
\label{tab1}
\end {table}

\begin{figure*}[bt]
\includegraphics[scale=0.35]{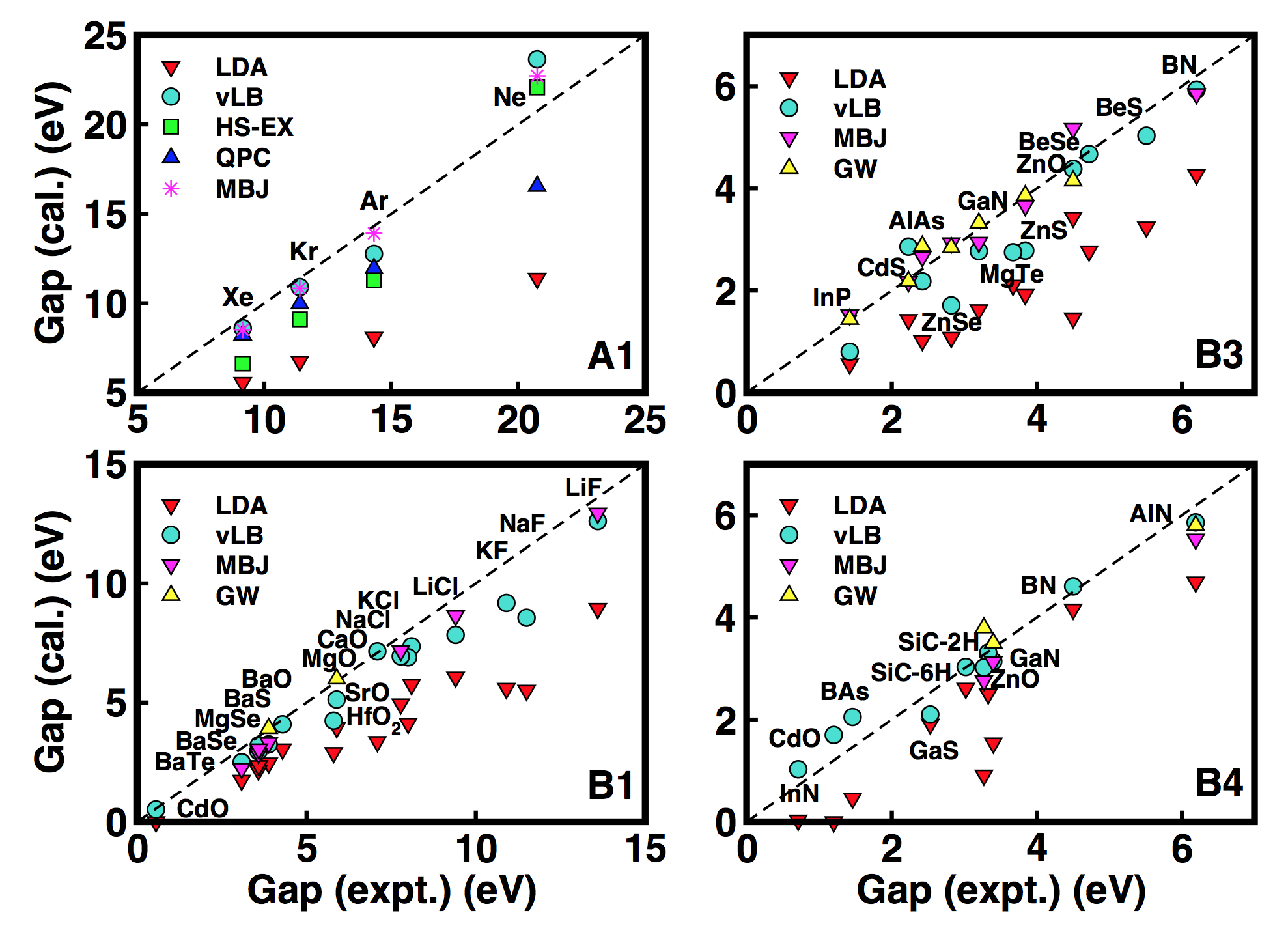}
\caption{(Color online).~Band gaps for materials with four structures:   (Left) A1 (FCC) with $\beta=0.04-0.082$ and B1 (Rocksalt) with $\beta=0.04-0.08$,\cite{GWCdO186, CdO186, SrO_GW} and (right) B3 (Zincblende) with $\beta=0.03-0.075$ and B4 (Wurtzite) with $\beta=0.03-0.09$.\cite{SiC_GW}}
\label{fig1}
\end{figure*}

{\par} Clearly, adding vLB-correction to LDA exchange improves the asymptotic behavior over LDA, and, if  used with the optimized-$\beta$ approach, it satisfies the IP-theorem for atom and diatoms due to its exact density description. Although all quantities used in Eq.~\ref{VLB} are semi-local, still they lead to a good approximation to the EXX-type of potentials because of correct treatment at shell limits, i.e., r$\rightarrow$0 to r$\rightarrow$R$_{ASA}$. This approach produces good band gaps of semiconductors and insulators which compare well with experiments,\cite{expt4,expt7,expt10,expt14,expt18,expt23,expt24,expt25,expt26,ExptRG,GaS} see Table~\ref{tab1} and Fig.~\ref{fig1}.

%

\subsection*{Wurtzite-ZnO}
{\par}A zinc oxide (ZnO) semiconductor remains a topic of interest because of its optoelectronic applications owing to its direct wide band gap $E_{g}\sim$3.40 eV at room temperature.\cite{75} The ZnO exists in wurtzite (B4), zinc blende (B3), and rocksalt (B1) crystal structures but in ambient conditions, the  B4 is thermodynamically most stable phase. After R\"ossler's prediction of Zn-3$d$ level at 12 eV below valence band maxima in B4-ZnO,\cite{76} several experiments\cite{77,78,79,80} were done and showed significant difference to the calculated result. Langer \etal \cite{77} and  Powell \etal\cite{78,79} used X-ray-induced photoemission spectroscopy and UV photoemission measurements, respectively, to determine the position of Zn-3$d$ core level and placed it at 7.5 $\pm$0.2 eV from valence band maxima -- $3~e$V lower than that predicted by R\"ossler. X-ray photoemission found similar values to UV, i.e., Vesely \etal\cite{81} at 8.5 eV and of Ley \etal\cite{80} at 8.81 eV.\cite{ZnOReview2005} Despite good agreement with qualitative valence-band dispersion from LDA functionals,\cite{82,83,84,85}  the debate on quantitative position of Zn-$3d$ level in B4-ZnO remains a good exercise for most semi-local functionals. So, we provide, as one test, results on B4-ZnO from the  optimized LDA+vLB potential within  TB-LMTO-ASA.

{\par} For ZnO, core and valence orbitals of (Zn, O) were set to ($1s2s2p3s3p3d$, $1s$) and ($4s4p3d$, $2s2p$), respectively. In calculation of band-energies, we use valence states of (Zn, O), i.e., ($4s4p3d$, $2s2p$) as the basis set. We have added two other lattices of empty spheres (ES1, ES2)  at [(0,0,0.34), ($-$0.29,0.5,0.249)] in the unit cell to obtain a close-packed structure to fulfill the criteria needed for the atomic-sphere approximation assumed in TB-LMTO-ASA.\cite{TBLMTO} Atomic sphere radii, $R_{ASA}$, of (Zn, O) has been fixed to (2.095, 1.775)$\AA$ in LDA+vLB, HS-EX and LDA calculations. The basis set used in calculation for [(Zn, O), (E1, E2)] is [($4s4p3d$, $2s2p$), ($1s2p3d$, $1s2p$)]  are complete under all symmetry operations and no additional basis atom has been introduced.

\begin{figure}[t]
\includegraphics[scale=0.34]{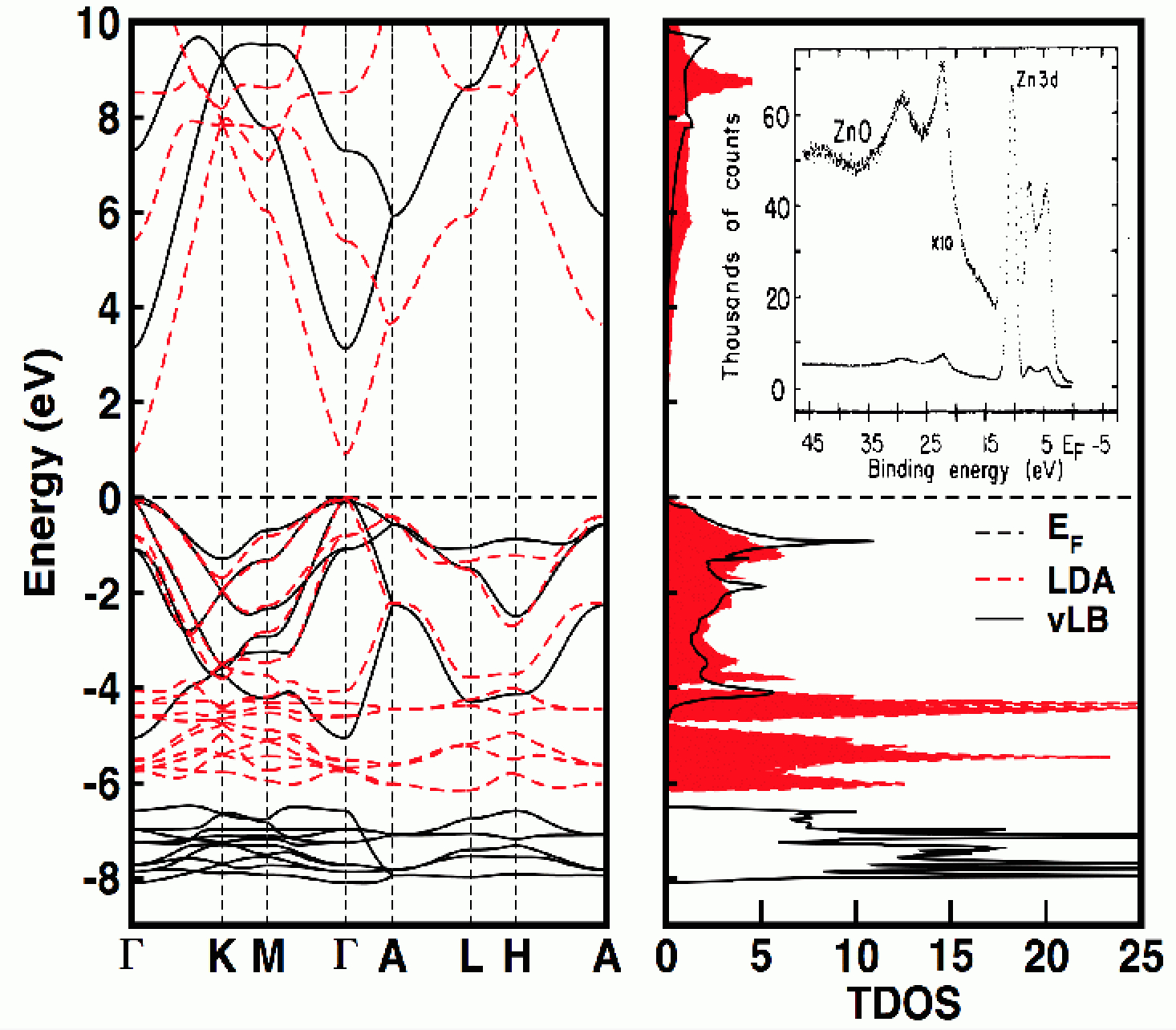}
\caption{(Color online).~The LDA+vLB results for wurtzite-ZnO gives Zn-3$d$ peak at 7.50$\pm$0.2 eV (LDA is at $\sim$6.0 eV). The X-ray photoelectron spectra of most stable polymorph, i.e.,  wurtzute-ZnO is given in inset.\cite{80}}
\label{fig2}
\end{figure}

{\par}For LDA, see Fig.~\ref{fig2}, we found Zn-3$d$ levels at $\sim6.0~e$V from Fermi energy E$_{F}$, meaning Zn-3$d$ levels are now incorrectly closer to O-2$p$ states, giving a stronger interaction with O-2$p$ levels. Increased interaction leads to strongly hybridized Zn-3$d$ and O-2$p$ states, which pushes O-2$p$ level towards the conduction band minimum, resulting into reduced band gaps. The typical error in estimating ZnO (B4) band gaps using LDA originated from strong Coulomb correlations between Zn-3$d$ and O-2$p$ levels. Although exact calculation of correlation is not possible, treating exact exchange is numerically possible. For optimized LDA+vLB, this is achieved by introducing vLB-correction to LDA exchange and tuning $\beta$ to optimal value through the IP-theorem using the atomic constituents. For ZnO solid, LDA+vLB gives Zn-3$d$ peak-center at $7.40~e$V, in  agreement with measurements $7.5\pm0.2~e$V. Clearly, eigen-energies of semi-core Zn-3$d$ levels (incorrect in LDA and underestimated with respect to experiment by $\sim$3 eV) are corrected by introduction of the vLB asymptotic correction introduced here. The band gap of wurtzite-ZnO calculated using optimized LDA+vLB is $\sim3.10~e$V, which compares well with  observed band gap of $3.4~e$V\cite{80} while LDA value is $\sim1.0~e$V (an underestimation of $\sim$71\%). The band gaps of wurtzite-ZnO calculated from optimized-vLB versus other methods (MBJLDA,\cite{TB2009} HSE06,\cite{OTTPK2008} G0W0,\cite{SrO_GW} GW\cite{expt18})  are $\sim3.10~e$V and (2.68,\cite{TB2009} 2.49,\cite{OTTPK2008} 2.51,\cite{SrO_GW} 3.80\cite{expt18}) $e$V, respectively. Only the optimized-vLB (10\% too small) and GW (10\% too large) results are in reasonable agreement with experimental values.

\subsection*{Half-Heusler alloys}
{\par} {\bf Non-spin-polarized compounds:}~In this section, we revisit the work of Kieven \etal \cite{KKNFG2010} on  I$-$II$-$V (eight electron) half$-$Heusler systems that have prime importance in optoelectonics. The half$-$Heusler structure basically arises from three interpenetrating fcc lattices of X, Y and Z atoms crystallized in ternary XYZ compounds with $F{\overline{4}}3m$ space group. The atoms X, Y and  Z are arranged at positions $(1/2,1/2,1/2)$, $(0,0,0)$ and $(1/4,1/4,1/4)$ in units of cubic lattice parameter and can be viewed as a zinc$-$blende$-$like structure. The strongly bound valence electrons in I$-$II$-$V half-Heusler compounds separates the conduction and the valence bands resulting in a semiconducting behaviour with varying band gaps.\cite{KFS2006}  We considered 18 XYZ (X=Li, Na, and K, Y =Mg, Ca, and Zn, and Z=N and P; X, Y and Z belong to the first (I$-$A), second (II$-$A and II$-$B) and fifth (V$-$A) main group (subgroup) of the periodic system of elements) compounds with the half-Heusler structure and calculated band gap using TB-LMTO-ASA with LDA+vLB potential, in most cases, we find a good agreement with Kieven \etal\cite{KKNFG2010} and experiments.\cite{KK1988, KKT1994, KK1998, KNK2002}

\begin {table}[b]
\caption{Band gaps of I-II-V (B3) half-Heuslers: LDA+vLB  with $\beta=0.05$ (ours),  compared to experiment, LDA (ours), as well as GGA, B3LYP, and GW.\cite{KKNFG2010, KK1988, KKT1994, KK1998, KNK2002} All calculations use lattice constants from Kieven \etal \cite{KKNFG2010}}
\begin {tabular}{ccccccccccccc}\hline \hline
System&&& \multicolumn{7}{c}{Band Gap (eV)}\\ 
               &&  Expt. &&  vLB   && LDA  &&  GGA   && B3LYP && GW \\ \hline        
LiMgN     && 3.20  &&  3.37 && 2.85 &&  2.29  && 4.37   &&   ---    \\
LiMgP     && 2.43  && 2.07  && 1.93  &&  1.55  && 2.90   &&   ---   \\
LiCaN     &&  ---    &&  3.71  && 2.38  &&  2.21  && 3.78   &&   ---    \\
LiCaP     &&  ---    &&  2.91  && 2.23  &&  1.95  &&  ---      && 2.93 \\
LiZnN     && 1.91 &&  1.67  && 0.78 &&  0.52  && 2.34   &&   ---  \\
LiZnP     && 2.04 &&  1.17  && 1.32 &&  1.35  && 2.66   &&   ---   \\
NaMgN   &&  ---   &&  2.72  && 1.06 &&  0.77  &&  2.08 &&  ---    \\
NaMgP   &&  ---   && 2.23   && 1.54 &&  1.47  &&  2.76 && 2.79 \\
NaCaN   &&  ---   && 1.82   && 1.47 &&  1.15  &&  3.03 &&  ---   \\
NaCaP   && ---    && 1.74   && 1.01 &&  1.95  &&   ---    && 2.95  \\
NaZnN   && ---   && 0.06    && 0.00 &&  0.00  &&   ---   && 1.83   \\
NaZnP   && ---   && 0.00    && 0.30 &&  0.44  &&  1.64 && ---    \\
KMgN    && ---   && 1.05    && 0.33  &&  0.13  &&  ---    &&  ---      \\
KMgP    && ---   && 1.25    &&  0.97 &&  0.96  &&  ---    &&  ---     \\
KCaN    && ---   && 2.24    &&  0.82 &&  0.68  &&  2.14 &&  ---    \\
KCaP    && ---   && 2.08    &&  1.56 &&  1.54  &&   ---    && 2.90 \\
KZnN    && ---   && 0.14    &&  0.00 &&  0.00  &&   ---    && 1.98  \\
KZnP    && ---   && 0.00    &&  0.00 &&  0.00  &&   ---    &&  ---   \\
\hline \hline
\end {tabular}
\label{tab2}
\end {table}

{\par} {\bf Spin-polarized FeMnSb:}~The spin-resolved band structure of the half-metallic compounds shows unusual properties: FeMnSb is one example with half-Heusler crystal structure discussed thoroughly by de Groot \etal \cite{GMEB1983} and Chioncel \etal \cite{CKAL2006} To showcase spin-polarized vLB-corrections, we  consider FeMnSb. The standard representation of FeMnSb with C1$_{b}$ crystal structure contains three atoms: Fe(0, 0, 0), Mn(1/4, 1/4, 1/4), Sb(3/4, 3/4, 3/4) and a vacant site at (1/2, 1/2, 1/2) (replaced by chargeless ES for ease in calculations), respectively. In FeMnSb, Fe ($-$1$~\mu_{B}$) and Mn (3~$\mu_{B}$) moments stabilizes the gap and the half-metallic electronic structure with ferrimagnetic coupling with total moment of 2~$\mu_{B}$. The integer spin moment per unit cell criteria is one of the requirements for half-metallicity. 

\begin{figure}[t]
\includegraphics[scale=0.25]{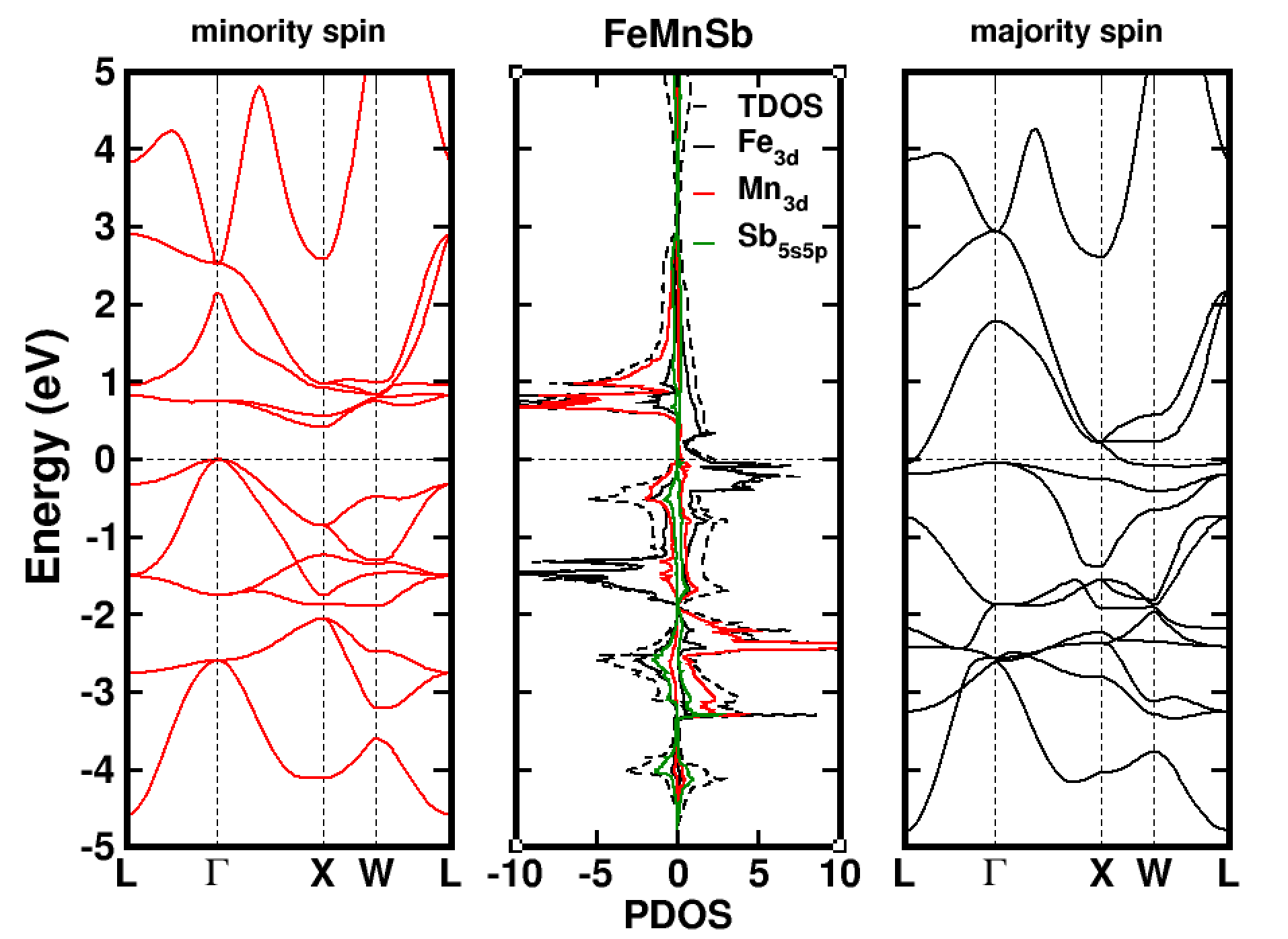}
\caption{(Color online).~Spin resolved band structure of half-Heusler half-Metallic FeMnSb alloy calculated with LDA+vLB potential at lattice constant of 5.703 $\angstrom$.\cite{GMEB1983,CKAL2006}}
\label{fig3}
\end{figure}

{\par} In Fig.~\ref{fig3}, electronic states for majority-spin projection have a metallic character with a nonzero density of states at E$_{F}$, the states with the minority-spin projection demonstrate a band gap at E$_{F}$.\cite{GMEB1983, IK1994} The gap originates from strong hybridization between the 3$d$-states of the transition metals Fe and Mn. This hybridization results in fully bonding states in the valence band and empty antibonding states in the conduction band, leading to an finite gap at E$_{F}$ (marked by a dashed line at zero energy). The deep lying $sp$-states of Sb do not have much effect on density of states at E$_{F}$, so it is not responsible for the existence of the minority gap. As a result, half-metal can, in principle, conduct a fully spin-polarized current, and hence it attracts attention for potential spintronics applications.\cite{IK1994, ZFS2004} 

\subsection*{Quantum dots: boron-nitride and graphene}
{\par} Two-dimensional materials, like of graphene and hexagonal boron nitride (h-BN), have drawn tremendous attention in terms of both  fundamental physics and possible applications in energy-generation devices.\cite{NGPN2009,GN2007,GHTMTZ2010} Single layers of graphene and h-BN have been fabricated and found to be stable at room temperature.\cite{NJSBKMG2005,MGKNBR2006,GBBWNG2008,JLSI2009,NRHDWR2010} The electrical conductivity in both cases vary largely because graphene is a semimetal and a very good conductor, while BN is an insulator (band gap$\sim${6 eV}).\cite{W1947, BRLC1995} 

{\par} We modeled one (3-B, 3-N or 6-C atoms), three (6-B, 7-N or 13-C atoms), and seven (12-B, 12-N or 24-C atoms) ring size quantum dots of boron-nitride and graphene within a cell of orthorhombic symmetry with cell parameters a$\eq$c$\eq$40$\angstrom$ and b$\eq$25$\angstrom$. The experimental geometry is used to generate 2D quantum dots.\cite{CK1982, SWE1995}

{\par} We calculated the LDA and LDA+vLB gaps between HO molecular orbitals (HOMO) and LU molecular orbitals (LUMO) of boron-nitride dots, where  boron and nitrogen have (3,3), (6,7) and (12,12) atoms each in the basis sets. Because TB-LMTO uses the ASA, we fill the rest of cell volume with ES and include them in basis set along with the atoms. As shown in Table~\ref{tab3}, with increasing dot size we approach the bulk optical band gap. The LDA+vLB potential within TB-LMTO-ASA yields a HOMO-LUMO gap of $5.70~e$V for B$_{12}$N$_{12}$ (Table~\ref{tab3}) that compares reasonably to the bulk band gaps observed  ($3.6~e$V\cite{F1971}  and $5.9~e$V\cite{B1972}) and  predicted (2.45 eV \cite{NS1971} to 5.4 eV \cite{DP1969}), while LDA largely underestimates all values.

\begin {table}[]
\caption{Band gaps of boron-nitride and graphene quantum dots (1, 3 and 7 ring sizes) calculated using LDA+vLB ($\beta=0.05$) and LDA potentials.}
\begin {tabular}{ccccc|ccc}\hline 
& \multicolumn{6}{c}{Band Gap (eV)}\\   \hline
ring-size & B-N & Expt & vLB  & LDA & C  & vLB & LDA \\ \hline
1 &B$_{3}$N$_{3}$ &---& 2.00  & 0.50 & C$_{6}$ & 2.90 & 1.50 \\
3 &B$_{6}$N$_{7}$ &---& 3.00  & 1.90 & C$_{13}$ &3.46 & 1.85 \\
7 &B$_{12}$N$_{12}$&---& 5.70  & 2.40 & C$_{24}$ & 5.20 & 2.50 \\
Bulk & h-BN & 3.60-5.90$\cite{F1971,B1972}$ & 4.60 & 3.90 & --- & --- & ---\\ 
\hline 
\end {tabular}
\label{tab3}
\end {table}

\section{Conclusion}
Since KS-DFT was first proposed, a search has remained unabated for a quality but numerically fast exchange-correlateion functional  to predict band gaps correctly. Here, we presented results using a spin-polarized vLB-corrected potential, which matched asymptotic behavior of exchange at the atomic sphere boundary (i.e., local interstitial in the solid) and which also satisfied the ionization potential (IP) theorem for atomic constituents. The combination approximately enforces the ionization energy and HOMO-LUMO difference to agree in first-principle calculations. The LDA+vLB-corrected exchange in combination with IP-theorem may be a good candidate to fill the gap of orbital-dependent functionals using semi-local quantities, and it provides an approximate exact-exchange band structure with no more computational cost that LDA or GGA. Compared with experiments, our asymptotically-corrected LDA obtains accurate band gaps for semiconductors and insulators, where in some cases it yields  gaps comparable to or better than more sophisticated XC methods, such as hybrid, exact-exchange, or GW.

\section{ACKNOWLEDGMENT}
We  thank A. Alam (IIT Mumbai, India) and W.A. Shelton (LSU) for critical comments.  Work at Ames Lab was supported by the U.S. Department of Energy (DOE),  Office of Science, Basic Energy Sciences, Materials Science and Engineering Division. The research was performed at the Ames Laboratory, which is operated for the U.S. DOE by Iowa State University under Contract No. DE-AC02-07CH11358. DDJ also acknowledges support from DOE for a CMSN grant through Brookhaven National Laboratory.

\end{document}